\documentclass[11pt]{article}
\pdfoutput=1
\usepackage{jheppub}
\usepackage{graphicx}
\usepackage{amssymb}
\usepackage{amsmath,amssymb}
\usepackage{slashed}
\usepackage{hyperref}
\usepackage{caption}
\usepackage{xcolor}
\usepackage{dsfont}
\usepackage{verbatim}
\usepackage{subfig}

\newcommand\beq{\begin{equation}}
\newcommand\eeq{\end{equation}}
\newcommand\be{\begin{equation}}
\newcommand\ee{\end{equation}}

\title{Massive Islands}

\preprint{\today}

\author[a]{Hao Geng}
\author[a]{, Andreas Karch}
\affiliation[a]{Department of Physics, University of Washington, Seattle, WA, 98195-1560, USA}

\abstract{We comment on the role of the graviton mass in recent calculations of the Page curve using holographic ideas. All reliable calculations of the Page curve in more than 2+1 spacetime dimensions have been performed in systems with massive gravitons. A crucial ingredient in these calculations is the formation of islands, regions that contribute to the entropy of degrees of freedom located elsewhere. While most often simply ignored, it is indeed true that mass of the graviton does not appear to significantly affect the calculations that appeared in the literature. We use the freedom to change the graviton mass to give an extremely simple model of analytically tractable island formation in general dimensions. We do however note that if one attempts to take the limit of zero graviton mass, any contribution from the islands disappears. This raises the question to what extent entanglement islands can play a role in standard massless gravity.
}

\begin{document}
\maketitle

\section{Introduction}

One of the most exciting developments in the last year in our understanding of quantum gravity has been the demonstration that the Page curve \cite{Page:1993wv}, the conjectured time evolution of the entanglement between an evaporating black hole and its Hawking radiation, in certain simplified models of quantum gravity can be rigorously calculated using ideas from holography. For a recent easy to read review of these developments see \cite{Almheiri:2020cfm}. In order to have a theory of quantum gravity that is under theoretical control one wants to study black holes in (asymptotically) anti-de Sitter (AdS) space so one has a non-perturbative definition of quantum gravity in terms of the AdS/CFT correspondence. Since AdS effectively acts as a box, large black holes in AdS do not evaporate. In order to get a tractable model of an evaporating black hole in AdS, the authors of \cite{Penington:2019npb,Almheiri:2019psf} impose transparent boundary conditions on AdS, allowing radiation reaching the boundary of AdS to couple to an external heat bath.

\begin{figure}[h]
\begin{centering}
\subfloat[Zero temperature. \label{bcft}]
{\includegraphics[scale=0.19]{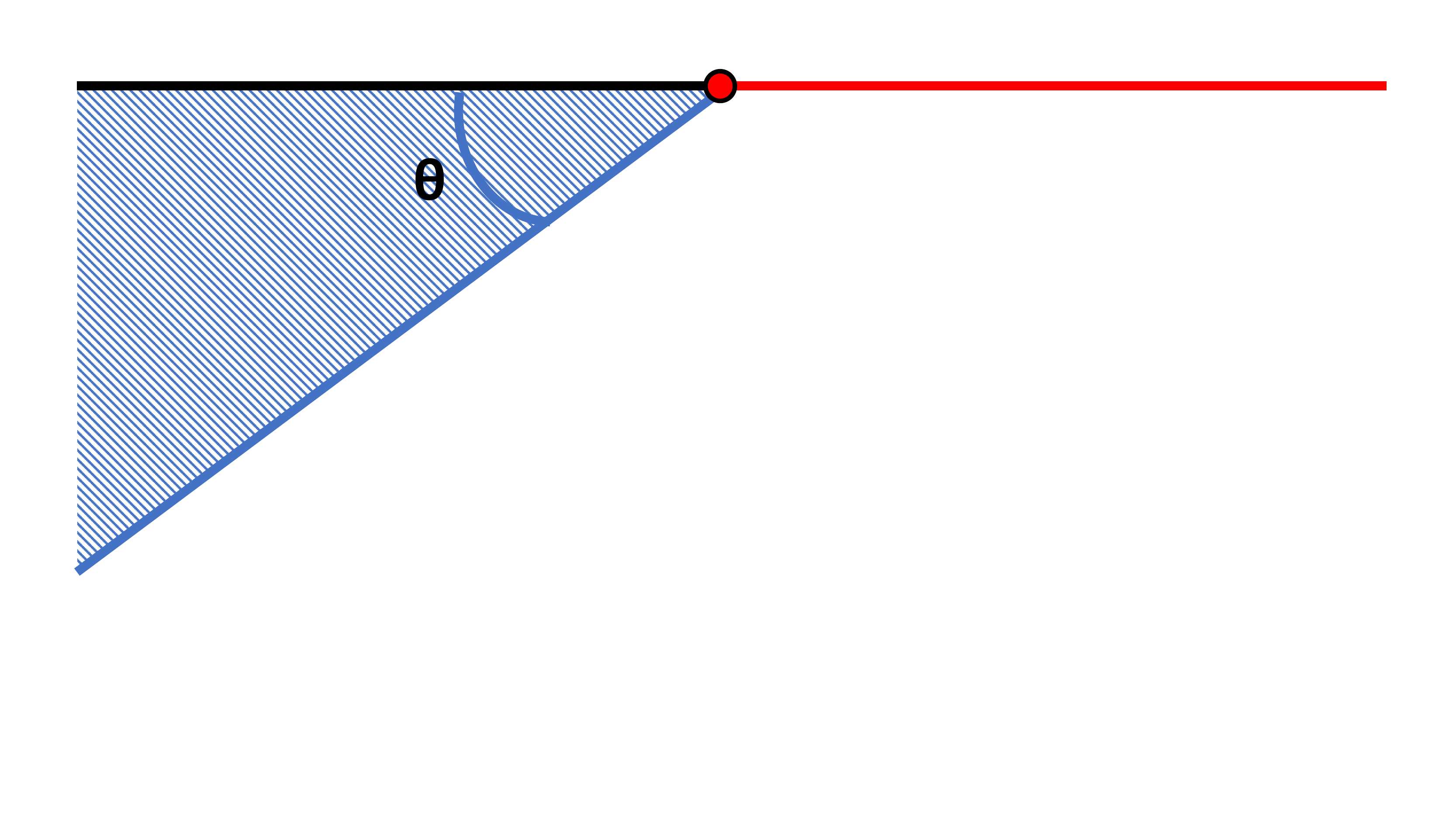}}
\hspace{1.5cm}
\subfloat[Finite temperature. \label{bcftblack}]
{\includegraphics[scale=0.19]{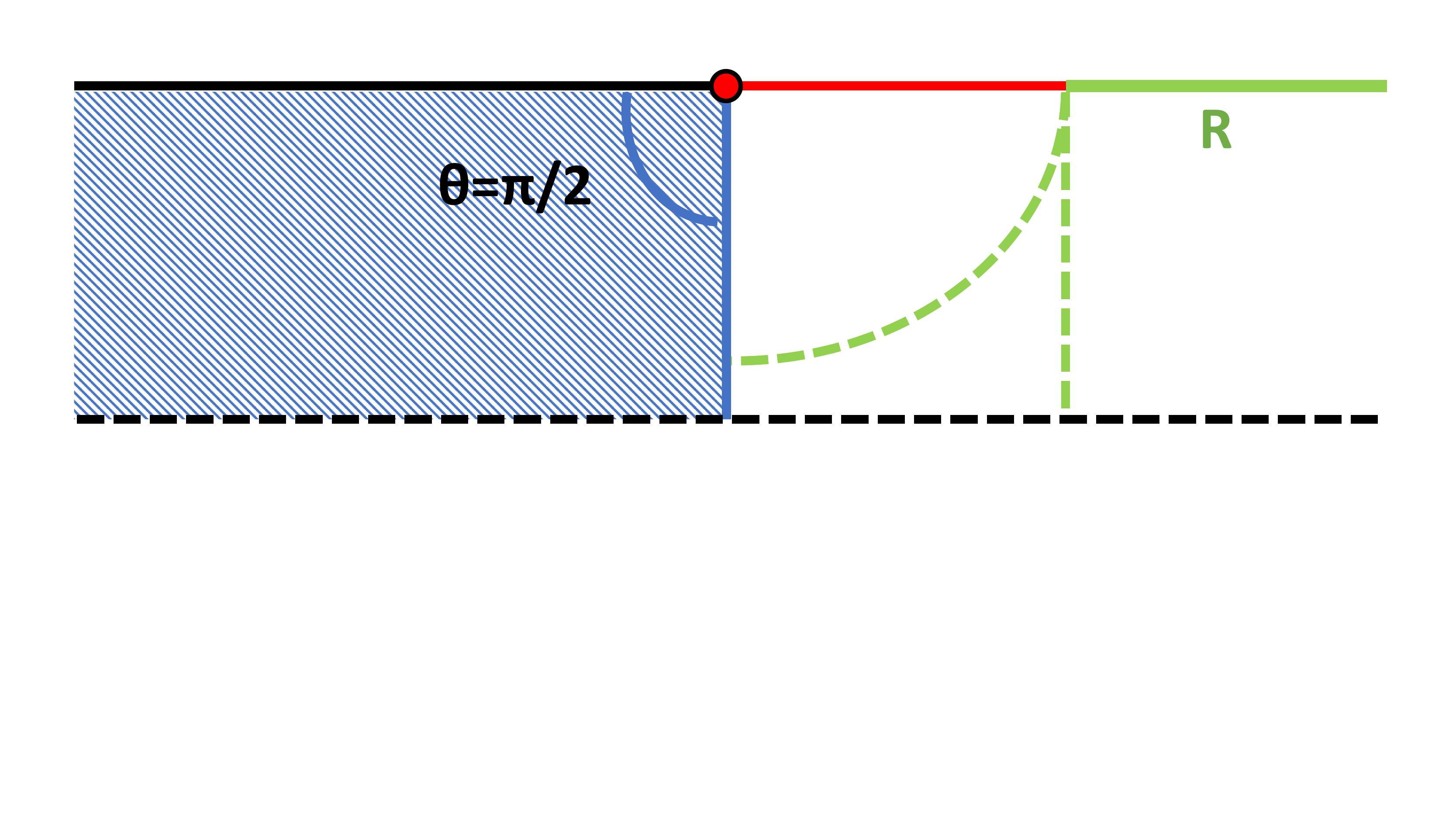}}
\caption{Embedding of subcritical RS branes in anti-de Sitter space and in the planar AdS Schwarzschild. R denotes the region whose entanglement entropy we wish to calculate, and the dashed green lines connected to the boundary of R are the two candidate RT surfaces. \label{bcfts}}
\end{centering}
\end{figure}

AdS with transparent boundary conditions has been explored extensively long before holographic calculations of the Page curves required to revisit this system. In \cite{Karch:2000ct,Karch:2000gx} it was argued, that a $d$ dimensional holographic field theory on a space with boundaries has three equivalent descriptions:
\begin{enumerate}
\item as a $d$ dimensional conformal field theory with a $d-1$ dimensional boundary (BCFT$_d$)
\item as gravity in an AdS$_{d+1}$ with an AdS$_d$ Randall-Sundrum  (RS) \cite{Randall:1999vf} brane
\item as a $d$ dimensional CFT living on the AdS$_d$ brane, coupled do a $d$-dimensional graviton, with transparent boundary conditions linking it to a CFT$_d$ on half space
\end{enumerate}
The first interpretation of RS branes in terms of BCFTs also has been emphasized and clarified in \cite{Takayanagi:2011zk}. It is the third ``semi-holographic" interpretation that brought up the consideration of transparent boundary conditions. It is also the description that is most useful for determining the Page curve. In order to understand this proposal, it is best to visualize the RS brane setup as displayed in figure \ref{bcft}.
This picture describes the holographic dual to a BCFT$_d$ at zero temperature, so the bulk is AdS$_{d+1}$ of curvature radius $R$. The RS brane cuts off parts of the bulk (the shaded part in figure \ref{bcft}). The embedding of the RS brane is given by the requirement \cite{Randall:1999vf} that the extrinsic curvature on the brane gets compensated by the brane tension\footnote{In the RS context one uses the brane to glue together two copies of the AdS spacetime and so the jump in extrinsic curvature is twice the extrinsic curvature of the submanifold occupied by the brane. To get to a BCFT one needs to orbifold to get a one-sided setup. In the the prescription of \cite{Takayanagi:2011zk} one jumps from a given value of extrinsic curvature to nothing, so the jump is only one times the extrinsic curvature. Besides a factor of 2 difference in formulas, this conceptual difference will come up again when discussing boundary conditions for RT surfaces below.}
\beq
\label{induced}
 K_{\mu \nu} - h_{\mu \nu} K = 8 \pi G T_{\mu \nu} = -4 \pi G \lambda h_{\mu \nu}. \eeq
Here $\mu$, $\nu$ run over the $d$ indices of the brane worldvolume with induced metric $h_{\mu \nu}$. $K_{\mu \nu}$ is the extrinsic curvature of the brane and $K$ is its trace. $T_{\mu \nu}$ is the stress tensor of the brane, whose action we take to be given purely by the tension times its volume. $\lambda$ hence is proportional to the tension of the brane. For a brane to have a timelike boundary as depicted in figure \ref{bcft} we need $|\lambda| < \lambda_c =(d-1)/(4 \pi G R)$. The wordvolume of the brane in this case has negative curvature and the induced metric itself is asymptotically AdS$_d$. It is only these subcritical branes with $|\lambda| < \lambda_c$ that are relevant in the context of BCFTs or transparent boundary conditions. When $|\lambda|$ passes $\lambda_c$ the boundary discontinuously changes to be lightlike or spacelike and the induced metric becomes Minkowski or de Sitter. For an extensive recent discussion of these scenarios see \cite{mismatched}.

The description 2) is in terms of d+1 dimensional gravity on the left-over AdS$_{d+1}$ space, that is the part not cut-off by the brane. One of the key insights of \cite{Randall:1999vf} was that this bulk description includes a $d$-dimensional mode of the graviton localized on the brane. The description 1) is the full holographic dual, a BCFT living on the part of the AdS$_{d+1}$ boundary that has not been removed by the brane, with boundary conditions imposed at the location of the intersection of brane and boundary. Last but not least, description 3) is the standard holographic interpretation of RS. We have a CFT living both on the half of the true boundary left behind as well as on the RS brane. The CFT on the RS brane has an intrinsic UV cutoff (corresponding to the removed region of spacetime behind the brane). It is also coupled to $d$-dimensional dynamical gravity -- the holographic manifestation of the RS graviton. As is obvious from the embedding as displayed in figure (\ref{bcft}), the CFT on the two halves of space (true boundary and RS space) are coupled to each other via transparent boundary conditions -- any excitations hitting the meeting point from the brane side gets transmitted to the true boundary side and vice versa.

One of the big surprises in the discussion of RS branes with subcritical tension ($|\lambda| < \lambda_c$), as relevant for AdS/BCFT or Page curves, was the realization that in this case (unlike the cases with $|\lambda| \geq \lambda_c$) the RS graviton always picks up a mass. This was first observed in \cite{Karch:2000ct} from considerations of bulk gravity, that is description 2). Soon after, Porrati demonstrated in \cite{Porrati:2001gx,Porrati:2001db} that in description 3) this mass is a direct consequence of the transparent boundary conditions. This is most clearly laid out in \cite{Porrati:2001db,Porrati:2002dt}: loops of matter fields on AdS$_d$ generically induce mass for the graviton, unless the matter fields obey standard Dirichlet boundary conditions. In particular, with transparent boundary conditions a graviton mass is always induced in complete agreement with the findings of \cite{Karch:2000ct} using the bulk description\footnote{A very nice quantitative check of this picture has been performed in \cite{Duff:2004wh} where it was shown that for ${\cal N}=4$ SYM a holographic calculation using the theory of AdS$_5$ with an RS brane (that is a description 1 calculation) gives the same graviton mass as a perturbative calculation in weakly coupled ${\cal N}=4$ SYM on AdS$_4$ with transparent boundary conditions (that is in description 3). While these calculations are valid in different parameter regimes, strong coupling for the first and weak coupling for the second, the results hint at a non-renormalization theorem. Certainly the numeric agreement is striking.}. The fact that transparent boundary conditions always induce a graviton mass is most obvious in the BCFT language of description 1), as emphasized in \cite{Aharony:2003qf}. Gravity on AdS$_d$ is dual to a theory living on the $d-1$ dimensional defect\footnote{We will keep on referring to the boundary of the BCFT as the defect, since the bulk spacetime of the BCFT itself is the boundary of AdS$_{d+1}$. We do not study defect CFTs in the sense that the field theory continues past the defect. The defect is genuinely the boundary of the BCFT. We are simply constrained by too many objects being called ``the boundary" in holographic duals of BCFTs.}  that is the boundary of the $d$ dimensional purely field theoretic BCFT description. In holographic dualities, the masslessness of the graviton is ensured by current conservation of the dual stress tensor. In a BCFT$_d$ the stress tensor living on the $d-1$ dimensional defect however is {\it not} conserved. Any boundary conditions that allow energy to be transferred from the AdS$_d$ brane (whose physics is encoded in localized degrees of freedom on the defect) to the bulk of the BCFT will lead to energy non-conservation on the defect and hence an anomalous dimension for the corresponding stress tensor and so a mass for the bulk graviton. Appearance of an anomalous dimension for the boundary stress tensor was verified by explicit field theory calculations in \cite{Aharony:2003qf}. This argument further makes it clear that massive gravitons are not limited to subcritical RS branes, but are a general consequence in {\it any} gravitational theory on AdS with transparent boundary conditions, as emphasized in \cite{Aharony:2006hz}. A gravitational theory on any AdS$_{d}$ space, even if it is not a worldvolume of an RS brane, has a dual description in terms of a CFT$_{d-1}$ with a conserved stress tensor. Transparent boundary conditions imply that we couple this CFT$_{d-1}$ (which for an RS brane is the theory living on the defect) to another system to which it can transfer energy. This immediately implies that the CFT stress tensor is not conserved and so can and hence will generate an anomalous dimension. The dual graviton gets a mass. If this coupling happens via a double trace deformation the graviton mass and the anomalous dimension of the stress tensor can be calculated independently on both sides and perfect agreement has been found \cite{Aharony:2006hz}, verifying the overall picture.

Since all recent calculations of the Page curve have been performed in systems with transparent boundary conditions, the corresponding holographic duals always have a non-conserved stress tensor and so, to the extent that they have gravitons at all, they all describe massive gravitons. Of course many of the explicit Page curve calculations have been done in low dimensional systems \cite{Almheiri:2019psf,Almheiri:2019yqk}. Neither AdS$_2$, where most of the work has been done, nor standard AdS$_3$ gravity support propagating gravitons. So in these cases it is meaningless to talk about the mass the graviton picks up due to the transparent boundary conditions. It is, however, still the case that the moment one allows for the dual CFT to couple to a bath the stress tensor of that dual CFT is no longer conserved and so will pick up an anomalous dimension.

Despite most of the recent activity on black holes being focused on low dimensions, it is believed that similar arguments also work in higher dimensions \cite{Penington:2019npb}. One of the few explicit calculations of black hole entanglement in this context has been performed by ref. \cite{Almheiri:2019psy} exactly in the setting of \cite{Karch:2000ct}, that is for a subcritical 4d RS brane in a 5d spacetime. The mass of the graviton, which is an unavoidable part of any theory of AdS gravity with transparent boundary conditions, has never been explicitly taken into account in any of these papers, even so implicitly its existence is being acknowledged but deemed not to matter. In this note we'll briefly remind the reader of some of the implications the graviton mass. In particular, we will analyze whether the graviton mass affects the existence of islands \cite{Almheiri:2019yqk,Almheiri:2019qdq,Penington:2019kki}, regions of spacetime that are completely disconnected and spacelike separated from a system of interest, but still partly determine the entanglement of degrees of freedom in that region of interest. We will show that going to an extreme limit in which the graviton is as heavy as it can get yields a very simple model of a higher dimensional black hole with entanglement islands. While very different from ``standard" AdS$_d$ gravity, we will show that this theory is not qualitatively different from the much more complicated constructions that have appeared in the literature already, most notably \cite{Almheiri:2019psy}. We will however also show that in the opposite limit, the physically most relevant case of zero graviton mass, the entanglement islands completely disappear.

This note is organized as follows. In the next section we will review some basic aspects of massive gravitons for subcritical RS branes. We'll recall that the mass of the graviton qualitatively changes as the curvature radius of the induced metric on the brane grows. In section 3 we'll review the importance of islands and the evidence for their importance in higher dimensions using subcritical RS branes leading up to our almost trivial model of a solvable brane world black hole with known holographic dual. In section 4 we confirm that the key properties of our black hole, as expected, still reproduce the effects first observed in \cite{Almheiri:2019psy}. In particular, we demonstrate the existence of islands that will dominate the entanglement entropy of ``the radiation" at late times. In section 5 we turn to the limit of the massless graviton. Most notably, we find that the contribution of islands to the entanglement entropy of the radiation vanishes in the limit of massless gravitons. We end with a brief discussion of what this finding may mean for the evolving story of holographic Page curves.

\section{Massive RS Gravity}

As we reviewed in the introduction, massive gravitons are unavoidable anytime we impose transparent boundary conditions on AdS gravity. For the rest of this work we want to focus on the particular realization of transparent boundary conditions in the context of RS gravity with its three dual representations reviewed above. Subcritical RS branes are controlled by a single real parameter, the ratio of brane tension to critical tension $x=\lambda/\lambda_c$. Geometrically, this parameter translates into the angle $\theta$ between brane and boundary as displayed in figure (\ref{bcft}). For positive tension branes, $\theta \leq \pi/2$. $\theta=\pi/2$ corresponds to the tensionless limit $x=0$. In this case the RS brane degenerates into a probe brane. It does not backreact on the geometry, its stress tensor vanishes. In the one-sided RS case we are currently considering it however still creates a boundary as we are still instructed to remove half of space. The brane is an orbifold plane\footnote{The idea that the simplest implementation of BCFTs is via orbifolds has recently been emphasized in \cite{Shashi:2020mkd}.}. As $x \rightarrow 1$ the angle $\theta \rightarrow 0$. In this limit several interesting things happen: a) Newton's constant $G$ goes to zero, b) the curvature radius on the brane diverges, c) the graviton mass goes to zero. It is important to understand this limit in more detail.

Let us collect a few important relations between the angle $\theta$, the ratio of tension to critical tension $x$, as well as the curvature radius $l$ of the AdS$_d$ space time on the RS brane:
\beq
\label{trig}
l^2 = \frac{R^2}{\sin^2 \theta} \approx \frac{R^2}{\theta^2}, \quad \frac{R^2}{l^2} = 1 - x^2, \quad x^2 = \cos \theta \approx 1 - \theta^2. \eeq
These relations follow directly from applying (\ref{induced}) to the setup of an AdS$_d$ RS brane in AdS$_{d+1}$.
The approximate expressions are valid at small $\theta$.
Note that, as expected, the brane becomes flat as we approach the critical tension from below. For the tensionless probe brane at $\theta=\pi/2$ the induced curvature radius is identical to the background curvature radius $R$.

Next let us take a look at $G^{-1} \sim M_d^{d-2}$ where $M_d$ is the $d$-dimensional Planck scale. Usually in RS setups we can determine $M_d$ from $M_d^{d-2} \sim M_{d+1}^{d-1} V$ where $V$ is the volume of the internal dimension (where we sliced AdS$_{d+1}$ in AdS$_d$ slices). This doesn't work in the subcritical case since the volume is truly infinite. While the RS brane cuts off the integral on one side, the volume still picks up an infinite contribution from the integration near the part of the true boundary that has not been removed by the RS brane. This is another manifestation of the mass of the graviton. The true massless graviton is not normalizable because of this infinity. The wavefunction of the lightest graviton, the ``almost zero mode" of \cite{Karch:2000ct}, is normalizable and gives us
\beq \label{gn} G^{-1}\sim  M_d^{d-2} \sim \frac{1}{\theta^{d-2}}  \quad \Rightarrow \quad  M_d \sim \frac{1}{\theta} \eeq
just as one would find for the normalizable zero mode in standard RS setups with $|x| \geq 1$ \cite{Randall:1999vf}.

Next let us look at the mass of the graviton. Unlike its $|x| \geq 1$ cousins, the spectrum of gravitational fluctuations around a subcritical RS brane is discrete \cite{Karch:2000ct}. For generic $\theta$, where the trigonometric functions in (\ref{trig}) are of order 1 and hence the induced curvature radius scales as $l \sim R$, the mass of {\it all} gravitational modes, including the putative ``massless" or light graviton, is of order $1/R$. In this sense, we are really pushing the holographic interpretation of RS. The general claim of RS is that the tower of excited graviton KK modes can be dualized into a field theory living on the RS brane, while the zero mode has to be added by hand in the holographic interpretation as a graviton living on the brane. For generic $\theta$ we have no parametric separation in mass between the lightest KK mode (which becomes the dynamical graviton on the brane) and the excited KK modes (which dualize into the CFT). But all this shows us is that the quantum effects of the CFT strongly alter the properties of the graviton (give it a mass of order the AdS curvature scale). The theory on the RS brane is a theory of gravity, but we would not necessarily recognize it as such since quantum effects from the matter sector strongly modify the properties of the gravitational theory. Most notably, black holes in such a theory are not expected to look like standard Schwarzschild black holes.

Last but not least, we have a comment on the UV cut-off. This is an intrinsic part of the holographic interpretation of RS setups and probably the least rigorous aspect of this proposal. What is the exact nature of the cutoff? Maybe recent work on the $T\bar{T}$ deformation \cite{McGough:2016lol} tells us that this is how we should interpret it. Just looking at the geometry, it is clear that the UV cutoff is taken to infinity in the $\theta \rightarrow 0$ limit. At $\theta=\pi/2$ the cutoff scale becomes of order the AdS curvature scale. So in this theory, the graviton mass is of order the cutoff scale, both of which are at the same scale as the curvature of the background geometry -- a very peculiar theory of quantum gravity indeed. At finite $\theta$ these scales separate by order 1 factors, but unless we take $\theta \ll 1$ there is no parametric separation of these 3 scales.

So lo and behold the only case in which a nearly massless graviton re-emerges is the $\theta \rightarrow 0$ limit. In this case, the KK tower is still composed of states whose mass goes as $m_{KK} \sim 1/l \sim \theta$. The lightest mode, the almost zero mode, in this limit however can numerically be seen to be much lighter \cite{Karch:2000ct}. A careful study reveals \cite{Miemiec:2000eq,Schwartz:2000ip} that it scales as $m_{AZM} \sim \theta^2$. This scaling is very important in the resolution of a puzzle that seems to immediately come to mind: as we take $x$ all the way to $x=1$ we should find a standard massless graviton. How can a massive graviton smoothly go over into a massless one? Even if the dispersion relation matches, at least in flat space the massive graviton propagates extra polarizations. This is known as the VvDZ discontinuity \cite{vanDam:1970vg,Zakharov:1970cc}. In AdS, all modes of the graviton pick up energies of order the AdS scale and so it is less obvious that there is a discontinuity as one takes the limit of zero mass. In fact, it was shown by Porrati in \cite{Porrati:2000cp} that there is no VvDZ discontinuity in AdS and, moreover, the double limit of taking the mass to zero together with the cosmological constant $1/l^2$ yields a massive flat space graviton if the mass scales as $1/l$, but a massless one if it scales as $1/l^2$ just like the almost zero mode does. Only in this limit does the lightest KK mode reveal its true character as a standard graviton.

The BCFT perspective, description 1), provides an alternative way to understand these properties of the graviton as was mapped out in detail in \cite{Aharony:2003qf}. In any BCFT, there exists a second operator expansion besides the OPE: the BOPE. Every operator of the d-dimensional ambient space\footnote{Following \cite{Aharony:2003qf} we will henceforth refer to the bulk of the BCFT as the ambient space, to keep the distinction from the $d+1$ dimensional bulk of the gravitational description 2).} can be decomposed into a series of operators living on the $d-1$ dimensional defect. Of most interest to us is the boundary operator expansion of the stress tensor. The $d$-dimensional stress tensor is conserved and hence has scaling dimension $\Delta=d$. It's boundary operator expansion yields a tower of operators living on the defect. As shown in \cite{Aharony:2003qf} their scaling dimensions $\Delta_n$ map 1-to-1 to the masses in the mode decomposition of the bulk graviton in description 2). Finding an almost zero mode would correspond to finding an operator in the BOPE of the stress tensor of dimension $d-1$. This requires at least one of the operators to acquire a larger anomalous dimension.

To understand why having an operator of dimension $d-1$ in the BOPE requires a large anomalous dimension, it is useful to zoom in on a solvable limit of the BCFT: the dual of $\lambda=0$, the probe brane limit. Recall that this corresponds to $\theta=\pi/2$ in the RS description. In this limit, there basically is no brane. How to interpret this limit, depends on whether one wants to take the two-sided bulk or a one-sided bulk which we get after imposing an orbifold constraint. Both cases have been considered in the RS literature. The two-sided bulk means we have two copies of the geometry displayed in figure \ref{bcft} glued together at the brane. In the BCFT language a two-sided geometry is dual to a defect rather than a boundary. In this two-sided case the $\lambda=0$ case means we simply have no boundary. We study a $d$-dimensional field theory living on Minkowski space and pick an otherwise undistinguished plane to call it the defect and apply the rules of BCFTs to it. In this case the BOPE is simply a Taylor series expansion and so the lowest operator appearing in the BOPE of the dimension $\Delta=d$ stress tensor also has dimension $\Delta=d$. All other operators appearing in the BOPE have dimension $d$ plus an integer. This spectrum is in perfect agreement with the bulk analysis of the $\lambda=0$ case. In the one-sided case that is of most interest to us here we are actually describing a true BCFT: we orbifold a known CFT on $d$ dimensional Minkowski space. In the bulk, this simply removes every other mode. It doesn't change the fact that the lowest dimension operator appearing in the BOPE has dimension $d$. The graviton is massive. Let us emphasize that this $\lambda=0$ limit is really special in that it is the {\it only} case in which we actually know what the dual CFT is. The RS brane is not a top-down object. While one can apply the standard rules of AdS/CFT to it, we do not know how to embed it as it stands into a string theory constructions. True stringy versions of RS are rather cumbersome and require extra ingredients. The orbifold is the only BCFT where we can start with a known AdS/CFT pair and get an RS setup for which we know precisely what the dual CFT is. This point, that orbifolds are the simplest ways to get AdS/BCFT pairs, has recently also been promoted in \cite{Shashi:2020mkd}.

The full 1-parameter family of RS systems corresponds to BCFTs where the anomalous dimension of the lowest operator in the BOPE interpolates from $d$ (at $\theta=\pi/2$) to $d-1$ (as $\theta \rightarrow 0$). Again, we see that at generic $\theta$ the graviton is massive. Only at $\theta=0$ do we recover standard gravity. All Page curve calculations and island constructions in dimensions where gravitons actually exist have been done in such a theory. As we will discuss below, this does not appear to be a problem. This is still a theory of quantum gravity with a reasonable Page curve. One point we wish to make here is that if we do calculations with a graviton whose mass is of order $1/l$ in any case (that is, a stress tensor whose BOPE starts with an operator of dimension $\Delta$ that is not infinitessimally close to $d-1)$, then one might as well go all the way to the theory at $\lambda=0$. The graviton is heavier by an order 1 factor over the ones considered in the literature, but the benefit is that one is working with a theory with known holographic dual. Furthermore, the bulk calculations are analytically tractable and do not require numerical GR.

\section{Islands and black holes}

One of the key ingredients in both the Page curve calculations of \cite{Penington:2019npb,Almheiri:2019psf} and the construction of entanglement islands in higher dimensions \cite{Almheiri:2019psy} is the use of a fully quantum version \cite{Engelhardt:2014gca} of the Ryu-Takayanagi (RT) entropy formula \cite{Ryu:2006bv}. For a gravitational theory in AdS, one can calculate entanglement entropies in the dual field theory by constructing a RT surface in the bulk whose boundary is the entangling surface in the dual CFT. The classical RT formula instructs us to find the minimal such surface; its area in Planck units gives the entanglement entropy. In the quantum version, instead of simply minimizing the area of a bulk surface one is instructed to also take into account the entanglement entropy of the bulk theory across the candidate surface. The true quantum RT surfaces minimizes the sum of area and matter entanglement terms. One key insight in the calculation of the Page curve in \cite{Penington:2019npb,Almheiri:2019psf} is that the true quantum entangling surface jumps in the process of black hole evaporation.

While this basic observation is believed to be very general, it is easiest to make this precise when we study the gravitational theory on an RS brane, that is use description 3) above. In this case, we can calculate the quantum contribution of the matter fields by using holography again, that is by going over to the bulk description 2). Since the bulk is classical, the standard RT surface in the AdS$_{d+1}$ bulk captures the quantum contribution of the CFT fields on the AdS$_d$ brane. Note that this necessarily requires the CFT living on the RS brane to have a large central charge. Alternatively, one can work in low dimensions, for example JT gravity in AdS$_2$ coupled to a CFT with large central charge \cite{Almheiri:2019psf}, in order to have control over the matter contribution.

In both cases, the quantum contribution to the entanglement entropy is dominated by the matter sector due to its large central charge $c$. In principle, fluctuating gravitons should also contribute. But this contribution is of order 1 in the large $c$ limit and so always subdominant. This is the main reason the graviton mass does not appear to play a major role in the story as it is told: the graviton contribution to the entanglement entropy is negligible to begin with.

The calculation of the Page curve is especially puzzling from the point of view of the outgoing radiation. In the RS setup and its various dual incarnations the outgoing radiation can be captured by calculating the entropy of a region R in the ambient space of the BCFT, far away from the defect. This region R is indicated in figure \ref{bcftblack}. Naively R has nothing to do with any quantum extremal surface living on the worldvolume of the RS brane. The claim of the island prescription is that in order to calculate the entanglement entropy of any specific region we need to allow contributions from islands that are located elsewhere. In particular, for the entanglement entropy of R we get contributions from an island located near the brane world black hole. To some extent, this is a very concrete implementation of at least the spirit of ER=EPR \cite{Maldacena:2013xja}, where the degrees of freedom near the horizon are redundant with the degrees of freedom in R.

Most of the work on islands has been done in lower dimensions \cite{Almheiri:2019yqk,Almheiri:2019qdq,Penington:2019kki}. In particular refs. \cite{Almheiri:2019qdq,Penington:2019kki} were able to demonstrate that in 2d JT gravity the island contribution arises from replica wormholes when one calculates the entanglement entropy using the replica trick. Much less is rigourously known in higher dimensions, even though it is believed that the story of islands is applicable more generally. Most notably, \cite{Almheiri:2019psy} was able to establish the existence of entanglement islands in higher dimensions exactly in the context of subcritical RS branes. This issue is important, as some questions have been raised regarding the proposed solution of the Page curve. For example, ref \cite{Bousso:2019ykv} argues that the resolution of the Page curve in terms of semi-classical geometry necessitates that the dual theory is an ensemble average. While this is indeed true for the examples involving 2d JT gravity, it is not believed to be true in the higher dimensional AdS/CFT examples. While \cite{Almheiri:2019psy} did not actually calculate a Page curve, they at least established the existence of islands for a 4d RS brane in AdS$_5$. We will build on this work by giving an explicit example of a known dual AdS/BCFT pair in which islands can be shown to exist. The dual field theory is an orbifold of ${\cal N}=4$ SYM and does not involve any ensemble averages.

Before we go there, let us describe in a little bit more detail what \cite{Almheiri:2019psy} has actually done. To use the RS setup to calculate the properties of an evaporating black hole one has to solve the coupled system of brane and Einstein gravity. This is in general not analytically tractable and numerical GR has to be employed. Solving the full time dependent problem is challenging. Ref. \cite{Almheiri:2019psy} decided to study a slightly simpler problem: instead of considering an evaporating black hole, they went back to consider a static situation by placing the ambient space of the BCFT at a finite temperature $T$ as well. As long as this $T$ matches the Hawking temperature of the black hole on the brane, the black hole can be in equilibrium with the bath. While not suited to calculate the Page curve of an evaporating black hole, this setup is sufficient to establish the existence of islands. The basic geometry of this setup is described in figure \ref{bcftblack}. The bulk has the causal structure a standard 5d black hole and the brane intersects the horizon. The metric itself however is very different from that of a 5d Schwarzschild black hole. Of course the geometry has to be extended past the horizon and the full geometry will include an identical second asymptotic region connected to the part we displayed via an ER bridge. For generic angle $\theta$ planar AdS Schwarzschild does not support a brane obeying the RS matching condition (\ref{induced}) and so one needs to change the bulk metric. While the basic features of the geometry still look like figure \ref{bcftblack}, the actual geometry has to be found by numerically determining a metric that is able to support an RS brane. This has been done for $\theta=0.343024$ and $\theta=1.47113$ in \cite{Almheiri:2019psy}. Note that neither of these angles is parametrically small, so in both cases the graviton mass will be of order the AdS$_4$ curvature radius on the brane.

One almost trivial thing to note is that at $\theta=\pi/2$ the situation dramatically simplifies. The geometry now literally is the one depicted in figure \ref{bcftblack}: an AdS$_{d+1}$ planar Schwarzschild black hole intersected by a brane at $y=0$, where $y$ is one of the Cartesian coordinates. The full $d+1$ dimensional metric is
\beq
\label{metric}
ds^2 = - h(r) dt^2 + \frac{dr^2}{h(r)} + r^2 (d \vec{x}^2 + dy^2 )
\eeq
with
\beq
h(r) = r^2 \left ( 1 - \frac{r_h^{d-2}}{r^{d-2}} \right ) .
\eeq
The induced metric on the brane is
\beq
ds^2 = - h(r) dt^2 + \frac{dr^2}{h(r)} + r^2 \, d \vec{x}^2
\eeq
with the same $h(r)$. This is indeed a $d$ dimensional black hole in that it has a horizon and a singularity. But the details are quite different from the standard Schwarzschild black hole in AdS$_d$. The fall-off of the blackening function $h(r)$ is still the one of the $d+1$ dimensional black hole. This tells us that the black hole is strongly modified by the quantum backreaction of the CFT.

One can compare and contrast this with the answers found in \cite{Almheiri:2019psy}. There they found that for the larger of their two angles the 4d metric differed significantly from the AdS$_4$ Schwarzschild solution, whereas for the smaller of the two angles the agreement was nearly perfect. Still, $\theta=0.343024$ is not a parametrically small angle and so the graviton is definitely still heavy in this case, it's mass going as $\theta^2$ when compared to the curvature radius. While the $\theta=\pi/2$ case may not appear to look like 4d gravity at all, our main point is that it is qualitatively ``no worse" than the cases considered in the literature: it's a graviton made massive by CFT loop effects, whose properties hence differ significantly from standard AdS$_4$ gravity. But it's a theory of gravity with black holes nevertheless.

As mentioned before, the other great advantage of the $\theta=\pi/2$ case is that this is the only example of an RS brane for which we actually know the full AdS/BCFT dual pair. In general RS branes are not top-down solutions of string theory as they stand. They ``stand in" for more complicated solutions that share some of the essential features. For $\theta=\pi/2$ the brane doesn't backreact at all and so really isn't present. All that is left is an orbifold prescription that instructs us to identify the $y<0$ part of the geometry with the $y>0$ part of the geometry. One can get such a geometry by starting with any known holographic dual pair, say ${\cal N}=4$ SYM and its AdS$_5$ $\times$ $S^5$ dual, and then perform an $y \rightarrow -y$ orbifold on the field theory side. This is our proposal for a simple holographic setting in which to study the appearance of entanglement islands.

Let us close this section with a discussion of a further issue that we have not yet touched upon as it does not play an important role for us, but it is something one should keep in mind whenever attempting to use RS branes to study quantum gravitational effects. Implicit in all this work using the holographic description of RS branes to calculate Page curves is the assumption that the quantum effects on the brane, that is in description 3), get geometrized by the classical bulk of description 2). This at first glance appears reasonable and just what AdS/CFT says. But, in fact, this is far from obvious. There has been a long debate on whether Hawking radiation on RS branes is geometrized.

Refs. \cite{Tanaka:2002rb,Emparan:2002px} were the first to point out that by the standard rules of AdS/CFT quantum effects, including the Hawking radiation, on the RS brane should be encoded in the classical bulk geometry. This made a stunning prediction: in the critical $x=1$ case, describing a flat Minkoswki RS brane, no time independent classical solution with a brane localized black hole should exist as flat space black holes are unstable to Hawking radiation. A possible way out was proposed in \cite{Fitzpatrick:2006cd}: as long as the bulk black hole only emits color neutral ``glueballs" as opposed to elementary CFT modes, it's Hawking radiation would have an energy of order 1 in the large number of colors $N$ counting that underlies AdS/CFT, and so would be invisible to the classical bulk geometry, which only encodes energies of order $N^2$. The black hole effectively drives the CFT into a confining phase. For the case of a flat brane this question was conclusively settled on the ``Hawking radiation not visible" side in \cite{Figueras:2011gd} where a stable brane world black hole solution was found numerically. As long as this is true, the program of using RS branes to study islands and black hole evaporation using classical geometry is doomed.

The full story is fortunately more complicated. Ref \cite{Hubeny:2009ru} found that, depending on details, there can be phases of holographic theories with black holes in which Hawking radiation is, in fact, visible at the classical level in the bulk. Unlike in the class of black holes that mimic the confining behavior of a truly brane localized black hole like the one constructed in \cite{Figueras:2011gd}, called droplets, these alternate scenarios are called funnels. They involve connected horizons in which the horizon of the brane world black hole smoothly merges with a bulk black hole horizon. They exhibit deconfined behavior and the quantum effects of Hawking radiation are geometrized. Fortunately the black hole solution depicted in figure \ref{bcftblack} and used in \cite{Almheiri:2019psy} is clearly a funnel. So we do expect the classical bulk to know about entanglement and islands. This issue would presumably become important if we actually would want to study a Page curve this way, as a brane localized black hole with the boundary CFT held at zero temperature appears to correspond to a droplet. A Page curve using RS setups has been calculated in \cite{Almheiri:2019yqk}, but that was using a 2d brane in a 3d bulk. This case is special in that {\it all} $d=2$ brane world black holes are funnels \cite{Hubeny:2009ru}.

\section{Analytic construction of islands for a ${\cal N}=4$ SYM based AdS/BCFT pair}

\subsection{Islands from equilibrium}

In the last section we proposed the simplest model of an AdS/CFT system in which we expect to find islands. The field theory is ${\cal N}=4$ SYM on $\mathbb{R}^{3,1}$ orbifolded by $y \rightarrow -y$. We want to consider the theory at a finite temperature $T$. This theory has two dual descriptions. One is in terms of the standard AdS$_5$ $\times$ $S^5$ geometry, also orbifolded by $y \rightarrow -y$. This is the bulk description 2) of the introduction. The second dual description is in terms of 4d gravity living on the (tensionless) AdS$_4$ RS brane located at $y=0$ coupled to ${\cal N}=4$ SYM itself, albeit with a UV cutoff (which is of order the AdS curvature radius). This is description 3) from the introduction. The graviton in this theory acquired a mass of order the AdS curvature radius as well. It's pushing the limits of holography to think of this as a ``theory of gravity", but a point we made many times above is that this scenario is qualitatively no different from the ones that have been considered in the literature. Pushing the cutoff up by an order 1 factor and the mass of the graviton down by an order 1 factor separates the scales between the two, but unless one has a parametrically large separation, which is only achieved in the $\theta \rightarrow 0$ limit, one is always stuck with a graviton mass of order the cutoff scale.

Let us demonstrate that this simple model exhibits islands just the same as its slightly more reasonable cousins constructed numerically in \cite{Almheiri:2019yqk}. To see this, let's first review how islands arose in that case. The configuration of figure \ref{bcftblack} describes, using the language of description 3), a black hole on the brane with Hawking temperature $T_h=T$ in equilibrium with the finite temperature CFT living on half of the true boundary of AdS$_5$. We want to calculate the entanglement entropy (EE) of a region R as indicated if figure \ref{bcftblack} located at large $y$, say $y \geq y_0$ for some fixed $y_0$. As the system is thermal, the full black hole geometry involves a second copy of the CFT behind the horizon\footnote{This second copy, associated with the termo field double, should not be confused with the second copy of AdS that would be part of a two-sided RS geometry. We study only one-sided RS geometries.} and we want to include a copy of the $y>y_0$ region in this second CFT as part of our region R as well.

In description 2) the bulk is completely classical and so we can use the standard RT minimal area prescription to determine the EE of region R. There are two classes of RT surfaces that can give us the entanglement. One, already present before introducing the RS brane, is the surface going straight down through the horizon, connecting via an ER bridge to the second boundary of the fully extended double sided black hole. This entangling surface has been well studied in the past \cite{Hartman:2013qma}. Its area is infinite, but this is just the standard UV divergence of the EE and comes from degrees of freedom localized near $y_0$. This infinity can easily be regulated by introducing a cutoff surface near the boundary. One very important aspect of this RT surface is that its area changes if we consider time evolution of the black hole state. In particular, the area of this trival RT surfaces running straight into the horizon grows linear with time together with the length of the ER bridge \cite{Hartman:2013qma}.

As pointed out in \cite{Almheiri:2019yqk}, the presence of the RS brane allows a second kind of RT surface, ending on the RS brane. This is also indicated in figure \ref{bcftblack}. By constructing this second surface numerically \cite{Almheiri:2019yqk} found that its area, at $t=0$, is in fact larger than that of the trivial surface. Since this non-trivial RT surface does not stretch across the ER bridge but ends on the brane outside the horizon its area however stays constant with time. That is, even though at time $t=0$ the trivial RT surface has the smaller area, the linear growth of the latter ensures that at some time $t_*$ the non-trivial surface takes over. But this non-trivial surface clearly corresponds to an island contribution! As is apparent in figure \ref{bcftblack}, the fact that this RT surfaces ends just outside the black hole horizon on the brane means that the degrees of freedom separated by the RT surface from the AdS$_d$ boundary on the brane constitute an island whose entropy ``counts" towards that of region R. At late times the entanglement entropy of the radiation system R is dominated by an island containing the braneworld black hole! Interesting enough, this same story can also be understood from the BCFT point of view, description 1), as recently demonstrated in \cite{Rozali:2019day}. There the competition is between different channels contributing to the correlation function of twist operators which calculate the EE. In any case, the entanglement island has been found from the non-trivial RT surface and dominates the late time behavior of the entanglement entropy.

To be precise, there exists in fact a one parameter family of non-trivial RT surfaces ending on the brane. For every location outside the horizon on the brane we can find a minimal area surface that connects it with the boundary of region R. Ref \cite{Almheiri:2019yqk} constructs all these surfaces and then picks the one with the minimal area. Instead of doing that, one could also have included a boundary term in variation of the area of the surface to begin with.\footnote{This was explicitly done after we published a previous version of this work in \cite{Chen:2020uac}.} We can also deduce the correct boundary condition the following way: Recall that the full RS geometry as it solves Einstein's equations has two identical copies of the geometry displayed in figure \ref{bcftblack}, including its analytic continuation. To reduce the full RS solution to a single sided copy we need to impose an orbifold projection across the brane. The RT surface needs to be symmetric in the double sided geometry to be consistent with this projection. This means it needs to end orthogonally on the RS brane. We will impose this boundary condition when constructing the RT surfaces in our new theory. We numerically verified that this boundary condition agrees with the procedure described in \cite{Almheiri:2019yqk} as well as the more recent discussion in \cite{Chen:2020uac}. The fact that the RT surfaces should end orthogonally on the RS brane has also been made from the point of view of the replica method in \cite{Jensen:2018rxu}.

\subsection{The minimal massive island model}

\begin{figure}[h]
\begin{centering}
\includegraphics[scale=0.59]{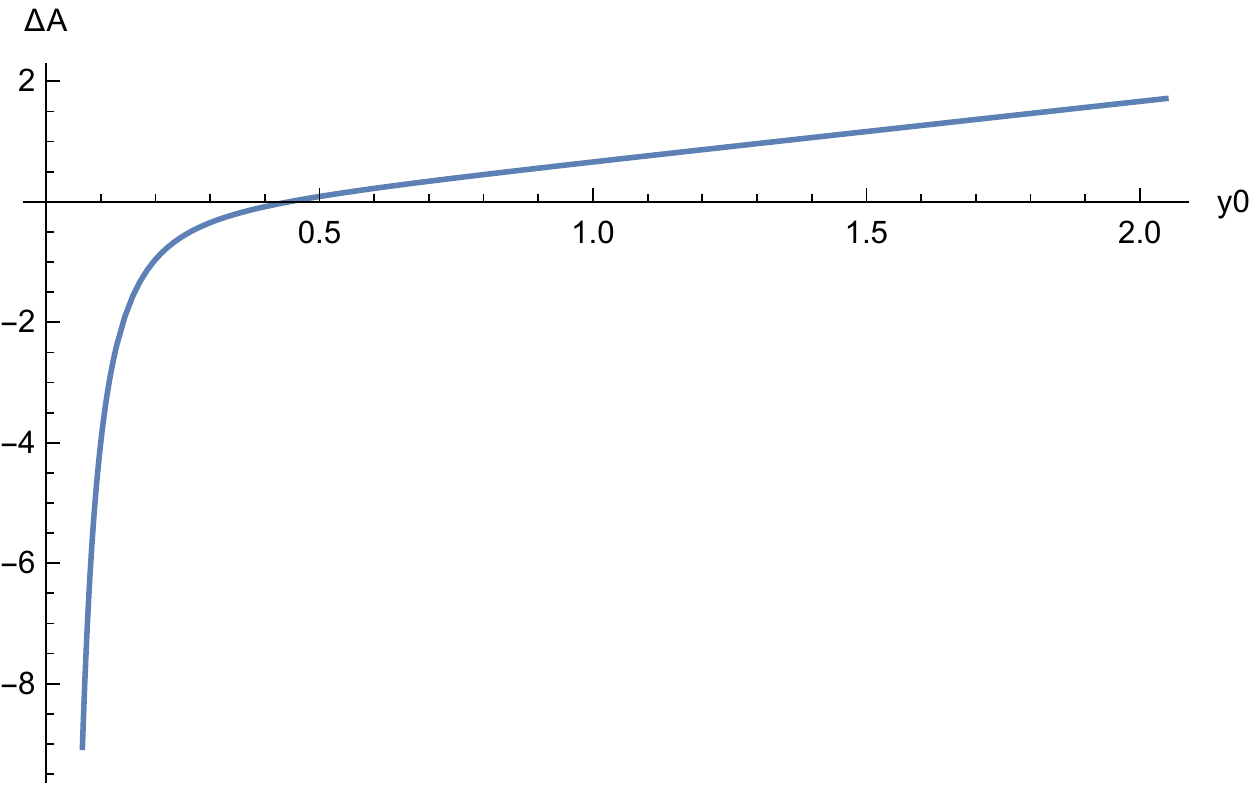}
\caption{Area difference $\Delta A$ as a function of $y_0$ in units where $z_h^{-1} = \pi T =1$.\label{areadifference}}
\end{centering}
\end{figure}

Let us revisit the issue of islands in the theory were boundary and black hole are in equilibrium in the simplest RS brane model, the one with $\theta=\pi/2$. The background metric is the one of (\ref{metric}). To keep the coordinates finite it is convenient to work with a new radial coordinate $z=1/r$. We also work in units where the curvature radius $R=1$. The metric now reads
\beq ds^2 = \frac{1}{z^2} \left ( - h(z) dt^2 + \frac{dz^2}{h(z)} + dy^2 + d \vec{x}^2 \right )
\eeq
with
\beq
h(z) = 1 - \frac{z^4}{z_h^4}.
\eeq
We are looking for a minimal area curve\footnote{We should note that the minimal areas in this geometry have also been constructed in a more general analysis of entanglement entropies in AdS/BCFT in \cite{Chang:2018pnb}.} with $t=0$ and $y=y(z)$ with the boundary condition that $y(0)=y_0$ and $1/y'(z_*)=0$ where $z_*$ is defined by the requirement $y(z_*)=0$. This latter boundary condition is easier to understand when writing the curve as $z(y)$. In this case we are saying that at $y=0$ we want $z'(0)=0$. As we argued before, this is required for the RT surface to be consistent with the $y \rightarrow -y$ symmetry by which we orbifold. In what follows we specialize to the case of $d=4$ (that is AdS$_5$ with an AdS$_4$ RS brane) to keep the formulas simple. Working out the general $d$ case is straightforward. The area per unit volume in $\vec{x}$ space of a general surface is given by
\beq
A = \int_0^{z_*} dz \, \frac{1}{z^3} \sqrt{\frac{1}{h} + (y')^2}.
\eeq
The EE is related to this via $S=A/(2G_5)$, where $G_5$ is the 5d Newton constant and the extra factor of 2 compared to the standard $A/(4G_5)$ is due to the fact that the full area in the complete analytically continued black hole geometry is twice of what we calculate in one asymptotic region.
Since $y$ doesn't appear in the action, we can easily find the solution for $y'$:
\beq (y')^2 = \frac{1}{h} \frac{(z/z_*)^6}{1- (z/z_*)^6}. \eeq
Here we fixed the integration constant by the boundary condition that $1/y'(z_*)=0$. We can write
\beq y(z) = \int_z^{z_*} dz \, \frac{1}{\sqrt{h}} \frac{(z/z_*)^3}{\sqrt{1-(z/z_*)^6}}. \eeq
This automatically obeys $y(z_*)=0$. $z_*$ is then related to $y_0$ by requiring $y(0)=y_0$.
Plugging back in the functional for the area density we find
\beq
A = \int_0^{z_*} dz \, \frac{1}{z^3 \sqrt{h} \sqrt{1-(z/z_*)^6}}.
\eeq
As expected, this area has a divergence near $z=0$. This is the standard UV divergence of the EE and corresponds to short-distance modes near the entangling surface at $y=y_0$. It would be easy to regulate this divergence. Instead of doing this explicitly, we calculate, as in \cite{Almheiri:2019yqk}, the difference in area between this non-trivial RT surface and the trivial RT surface given by $y=y_0$. This yields the finite area difference
\beq
\Delta A =  \int_0^{z_*} dz \, \frac{1}{z^3 \sqrt{h}} \left ( \frac{1}{ \sqrt{1-(z/z_*)^6}} -1 \right )
- \int_{z_*}^{z_h} dz \, \frac{1}{z^3 \sqrt{h}}.
\eeq
Switching to dimensionless variables $z = \xi z_h$, $z_* = \xi_* z_h$ this amounts to
\beq
(z_h^2) \Delta A = \int_0^{\xi_*} d \xi \, \frac{1}{\xi^3 \sqrt{1-\xi^4}}
\left ( \frac{1}{\sqrt{1- \xi^6/\xi_*^6}} - 1 \right ) - \int_{\xi_*}^1 d \xi \,
\frac{1}{\xi^3 \sqrt{1-\xi^4}}.
\eeq
The integrals can be done numerically. By doing so for various values of $\xi_*$ we can generate numerically pairs of $y_0$ and $\Delta A$. We plot the resulting $\Delta A$ as a function of $y_0$  in figure \ref{areadifference}.

Note that for large $y_0$ the area difference is positive. This is the region we are interested in. To take the region R as a stand-in for the escaping Hawking radiation we want it to be asymptotically far away from the defect. The scale for the $y_0$ at which we change from negative to positive area difference is set by the temperature $T$, as this is the only scale in the problem. Positive area difference is exactly what was found for the smaller angles in \cite{Almheiri:2019yqk} and one can repeat all the same arguments as were made above for that case. In summary, the island story carries over to our simple model unchanged.

\section{Massless Gravitons}

We have seen that theories with massive gravitons have entanglement islands in higher dimensions. We found a very simple model in which most calculations of this effect are analytically tractable. Last but not least, let us return to the question of what happens when we try to recover standard gravity. That is, we want to take the $\theta \rightarrow 0$ limit in which the graviton becomes massless. What is happening to the two RT surfaces in this limit? Clearly the trivial RT surface is unaffected by this limit. For large enough $y_0$ it is completely insensitive to the existence of the RS brane in the first place.

But interestingly enough, the non-trivial RT surface has an area that diverges as $1/\theta^{d-2}$. This is due to the fact that it now approaches the AdS$_{d+1}$ boundary on both ends. The corresponding divergence near $y=y_0$ is the UV divergence we encountered before, it's shared between both RT surfaces and cancels in the difference. But as the RS brane approaches the boundary as $\theta \rightarrow 0$ we now also get a UV divergences in the minimal area for the end near the brane. Holographically one can understand the $1/\theta^{d-2}$ divergences as arising from the vanishing $G$ on the brane. A finite area on the RS brane will give a contribution that diverges as $1/G$ when $G$ goes to zero. According to (\ref{gn}) this exactly accounts for the divergence of the non-trivial RT surface. Even though this explains the divergence, it is still true that the non-trivial RT surface will have an area that is parametrically larger than the trivial one. Even though the area of the trivial RT surface will grow linear in time, this means that the time $t_*$ at which its area exceeds that of the non-trivial one will also scale as $t_* \sim 1/\theta^{d-2}$. The island contribution never is important in the limit that the graviton mass goes to zero!

While we reached this conclusion in one particular class of models, RS branes and their holographic dual, it makes one wonder how general this phenomenon is. While islands are indeed important in theories of massive gravitons, the moment we try to take the mass of the graviton to zero we also lose the islands. As we reviewed, much of the work on islands and Page curves has been done in lower dimensions, most notably 1+1 dimensions. In this case, there is no graviton and so this question doesn't arise. It is however still true that in all these systems one has transparent boundary conditions and hence a non-conserved stress tensor in the field theory dual to the gravitational bulk. This is what forced the mass upon the graviton in higher dimensions. So one may wonder whether some of the similar issues arise in lower dimensions as well. In any case, it appears that if we try to move from toy models with massive gravitons to gravity that we would recognize the islands vanish, at least in examples where we have control.
However, it is important to point out that in the RS model, the disappearance of the islands happened since the limit of zero graviton mass was also the limit of vanishing $G$. And it is really the latter that drives the disappearance of the islands. So if one could divorce the two phenomena, one may be able to send the graviton mass to zero without losing the islands.
One could also wonder whether this phenomenon is related to the concerns of \cite{Bousso:2019ykv}, where it was argued that the determination of the Page curve consistent with unitarity literally is too good to be true -- semi-classical gravity should not have been enough. Of course it could also be the case that the issue of vanishing island contributions in the massless graviton limit is a peculiarity of using the RS brane model. Clearly more work is needed.

\section*{Acknowledgments}
This work was supported, in part, by the U.S.~Department of Energy under Grant No.~DE-SC0011637 and by a grant from the Simons Foundation (Grant 651440, AK). Special thanks to Ahmed Almheiri for patiently enduring our questions about the possible role of massive gravitons in the island story. We'd also like to thank Lisa Randall for comments and encouragement.

\bibliographystyle{JHEP}
\bibliography{islands}

\end{document}